\documentclass[conference]{IEEEtran}
\IEEEoverridecommandlockouts
\usepackage{cite}
\usepackage{amsmath,amssymb,amsfonts}
\usepackage{algorithmic}
\usepackage{graphicx}
\usepackage{textcomp}
\usepackage{xcolor}
\def\BibTeX{{\rm B\kern-.05em{\sc i\kern-.025em b}\kern-.08em
    T\kern-.1667em\lower.7ex\hbox{E}\kern-.125emX}}

\usepackage{algorithm}
\usepackage{algorithmic}
\begin{document}

\title{Adversarial Jamming for a More Effective Constellation Attack 
\thanks{*Corresponding author: Naijin Liu, (liunaijin@qxslab.cn) and Xueshuang Xiang (xiangxueshuang@qxslab.cn.).}}
 
\author{
\IEEEauthorblockN{Haidong Xie}
\IEEEauthorblockA{\text{Qian Xuesen Laboratory} \\
\text{China Academy of Space Technology}
}
\\
\IEEEauthorblockN{Nan Ji}
\IEEEauthorblockA{\text{Qian Xuesen Laboratory} \\
\text{China Academy of Space Technology}
}
\and
\IEEEauthorblockN{Yizhou Xu}
\IEEEauthorblockA{\text{School of Aerospace Engineering} \\
\text{Tsinghua University}
}
\\
\IEEEauthorblockN{Shuai Yuan}
\IEEEauthorblockA{\text{Qian Xuesen Laboratory} \\
\text{China Academy of Space Technology}
}
\\
\IEEEauthorblockN{Xueshuang Xiang*}
\IEEEauthorblockA{\text{Qian Xuesen Laboratory} \\
\text{China Academy of Space Technology}
}
\and
\IEEEauthorblockN{Yuanqing Chen}
\IEEEauthorblockA{\text{Qian Xuesen Laboratory} \\
\text{China Academy of Space Technology}
}
\\
\IEEEauthorblockN{Naijin Liu*}
\IEEEauthorblockA{\text{Qian Xuesen Laboratory} \\
\text{China Academy of Space Technology}
}
}

\maketitle

\begin{abstract}
  The common jamming mode in wireless communication is band barrage jamming, which is controllable and difficult to resist. Although this method is simple to implement, it is obviously not the best jamming waveform. Therefore, based on the idea of adversarial examples, we propose the adversarial jamming waveform, which can independently optimize and find the best jamming waveform. We attack QAM with adversarial jamming and find that the optimal jamming waveform is equivalent to the amplitude and phase between the nearest constellation points. Furthermore, by verifying the jamming performance on a hardware platform, it is shown that our method significantly improves the bit error rate compared to other methods.
\end{abstract}


\section{Introduction}

With the development of wireless communication technology, the corresponding jamming technology has become a popular research topic~\cite{1637931,5751298}. 
Wireless signals are allocated with different frequency bands, and different frequencies do not interfere with each other, which makes it possible to barrage regional targets. For example, near important meeting locations or college entrance examination halls, jammers can be used to barrage mobile phone frequencies and prevent illegal communication.

For a target frequency band, the most commonly used method is noise barrage. It is characterized by adding noise to a signal, and the amplitude distribution obeys a Gaussian distribution; this noise is commonly known as additive white Gaussian noise~(AWGN)~\cite{871393,1512123}.
This noise can mask the information contained in a signal. The larger the amount of noise is, the smaller the signal-to-noise ratio~(SNR) or signal-to-jamming ratio~(SJR) are and the worse the bit error rate~(BER) or symbol error rate~(SER) is, which causes poor communication performance.
When nothing is known about the transmitted signal characteristics, adding noise is almost the only choice. However, intuitively, AWGN is obviously not the best jamming waveform. Therefore, we are concerned about finding and constructing a waveform with the best barrage effect based on some signal feature knowledge.

An adversarial example~(AE) is a new concept recently proposed in the field of deep learning that aims to induce a model to output incorrect results that are similar to the real data~\cite{szegedy2013intriguing,2014arXiv1412.6572G}.
As deep learning technology has been successfully applied to the field of wireless communication~\cite{jagannath_machine_2019}, we believe that the signal receiving and demodulating process can be realized by a deep learning model, and the construction of a jamming waveform can correspond to the generation of AEs in deep learning.

Thus, we apply the idea of AEs to the jamming problem and propose the adversarial jamming~(AJ) waveform.
Our method first transforms communication signal demodulation and jamming into a deep learning model and then generates the corresponding AEs to obtain the jamming waveform, which can attack a demodulation model with minimum power.
In terms of effect, our method can optimize and find the best jamming waveform amplitude and phase, which can ensure that the information contained in the interfered signal is completely covered up so that the communication party cannot accurately obtain effective information.
When the constrained jamming power is low, our method can concentrate the jamming energy in a few waveforms to improve the jamming effect.

A numerical simulation and hardware verification show that our method has significant advantages over noise barrage.
The variation curve of SER with SJR shows that our method can achieve the best jamming attack under any condition. The hardware test shows that our method is fully practical, and the jamming effect will be further improved considering factors such as clock synchronization.

\section{Adversarial Jamming}


To use AE technology, we first need to transform the signal demodulation process into a deep learning process~\cite{de_vrieze_cooperative_2018}.
Formally, by constructing a dataset that includes waveform~$x$ and coding $y$ pairs,
demodulation becomes a classification problem.
Taking 16-quadrature amplitude modulation~(16QAM) as an example, coding $y$ correspond to $16$ different $4$-bit binary codes, and waveform~$x$ is a sinusoidal waveform with different amplitudes and phases. To enhance the anti-interference ability of the model, reference~\cite{de_vrieze_cooperative_2018} added a small amount of AWGN on the basis of a sinusoidal waveform.

\begin{figure}[tbp]{}
\vspace{-2mm}
\centerline{\includegraphics[width=\linewidth]{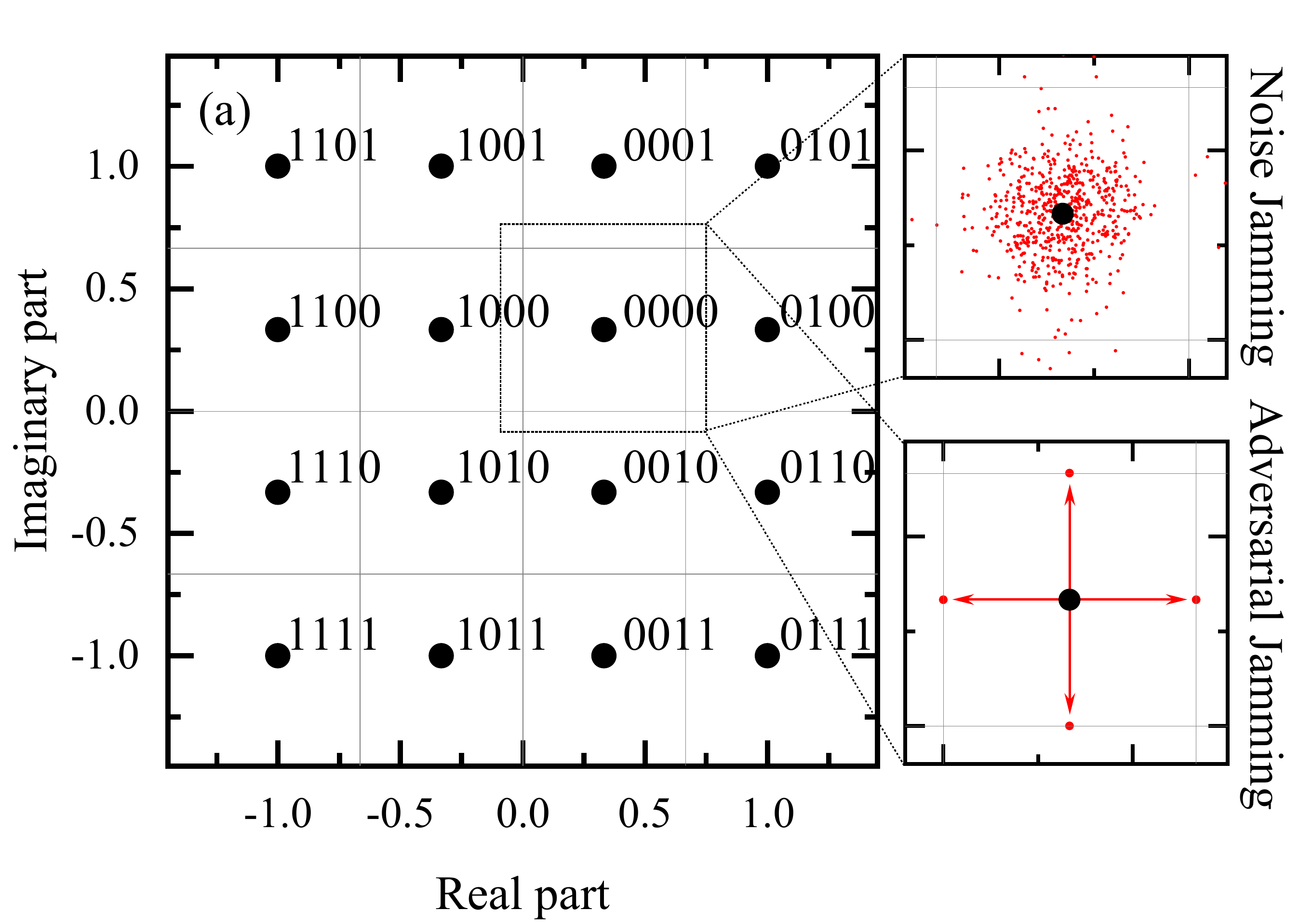}}
\vspace{-3mm}
\caption{(a) Constellation diagram of 16QAM (black dots), where thin black lines are used to mark the classification boundaries.
(b) Each red dot represents a constellation point position after AWGN jamming.
(c) Four red dots represent four possible positions of constellation points after AJ, which are randomly generated with the repetition of the algorithm. The red arrow represents the displacement vector, that is, the AJ waveform jamming effect.
\vspace{-5mm}
}
\label{fig:1}
\end{figure}

Therefore, we can train a classifier with excellent performance to realize the purpose of waveform demodulation; $y\!=\!f(x)$. Fig.~\ref{fig:1}(a) shows the constellation of 16QAM, in which the black dot corresponds to the amplitude and phase of the original waveform, the numbers near the dots correspond to their binary codes, and the red dispersed point cloud in Fig.~\ref{fig:1}(b) represents the influence of noise on the waveform.

The characteristic of AWGN is that the amplitude obeys a Gaussian distribution, that is,~$N(0, \sigma^2)$, and the phase obeys a uniform distribution, that is,~$U(0,2\pi)$.
Therefore, when the jamming power significantly exceeds the mean distance of constellation point $d$, the jamming waveform is sufficient to confuse and barrage the target. Otherwise, if it is small, most jamming waveforms are not sufficient to cause effective jamming, and the probability of an effective jamming waveform is $O(e^{-d^2/(2\sigma^2)})$. When the jamming power is medium, the improvement space of the Gaussian waveform is obvious. The random phase also limits the waveform jamming ability.

\textbf{Adversarial Jamming: }
In the form of AEs, the following problems are optimized: $\min{|\!| \delta |\!|_2}, \text{ s.t. } f(x+\delta) \neq y$, which mean the minimum perturbation $\delta$ that leads to an incorrect model result is found~\cite{szegedy2013intriguing}. When $f$ corresponds to a neural network, typical algorithms include FGSM~\cite{2014arXiv1412.6572G}, PGD~\cite{madry2017towards} and C\&W~\cite{cw}.
Thus, for the problem of attack or jamming demodulation, we can directly use the generation method of AEs to generate jamming, and
the waveform can be in the form of IQ data obtained by the receiver.

Due to the known constellation, we can directly relate the waveform to the coordinate points in the constellation. Therefore, finding the best jamming waveform essentially becomes the geometric problem of finding the nearest constellation point. After the numerical experiments, we found that the jamming waveform numerically solved by the AE method is often the displacement vector corresponding to the nearest constellation point, which is the minimum necessary for jamming with only a difference in numerical precision, as shown in Fig.~\ref{fig:1}(c).
If the desired SJR is low, i.e., the allowable jamming is stronger than the generated AEs, there is no need to change or enhance the perturbation amplitude to ensure the jamming effect.
On the other hand, if the allowable average jamming power is weaker than AEs, we cannot simply reduce the size of the perturbation because it will invalidate the jamming outcome. Here, our strategy is to use time intermittent jamming to ensure that a single jamming waveform still has enough perturbation, but the occurrence time of jamming decreases accordingly to meet the target average jamming power, $t\% \leftarrow P_{Jamming}/P_{Signal}$.
Finally, we summarize the AJ algorithm in Alg.~\ref{alg:train}.

\begin{algorithm}[t]
  \caption{Generating and Evaluating Adversarial Jamming}
  \label{alg:train}
  \begin{algorithmic}[1]
    \STATE {Generate a dataset with waveform $x$ and coding $y$ pairs,}
    \STATE {Train a demodulation model $f$ on the dataset,}
    \STATE {Generate AEs~(the jamming waveform) with less perturbation by an AE algorithm, such as PGD,}
    \IF {Signal power / Jamming (AEs) power $<$ SJR}
    \STATE {Increase the jamming waveform perturbation option,}
    \ELSE 
    \STATE {Reduce the occurrence time of the jamming waveform,}
	\ENDIF
    \STATE {Add the jamming waveform and original waveform to obtain the received waveform,}
    \STATE {Interpret the received waveform with model $f$, }
    \STATE {Count the SER of demodulation.}
  \end{algorithmic} 
\end{algorithm} 

\section{Experimental Result}

\subsection{Jamming Analysis}

\begin{figure}[tbp]
\centerline{\includegraphics[width=0.9\linewidth]{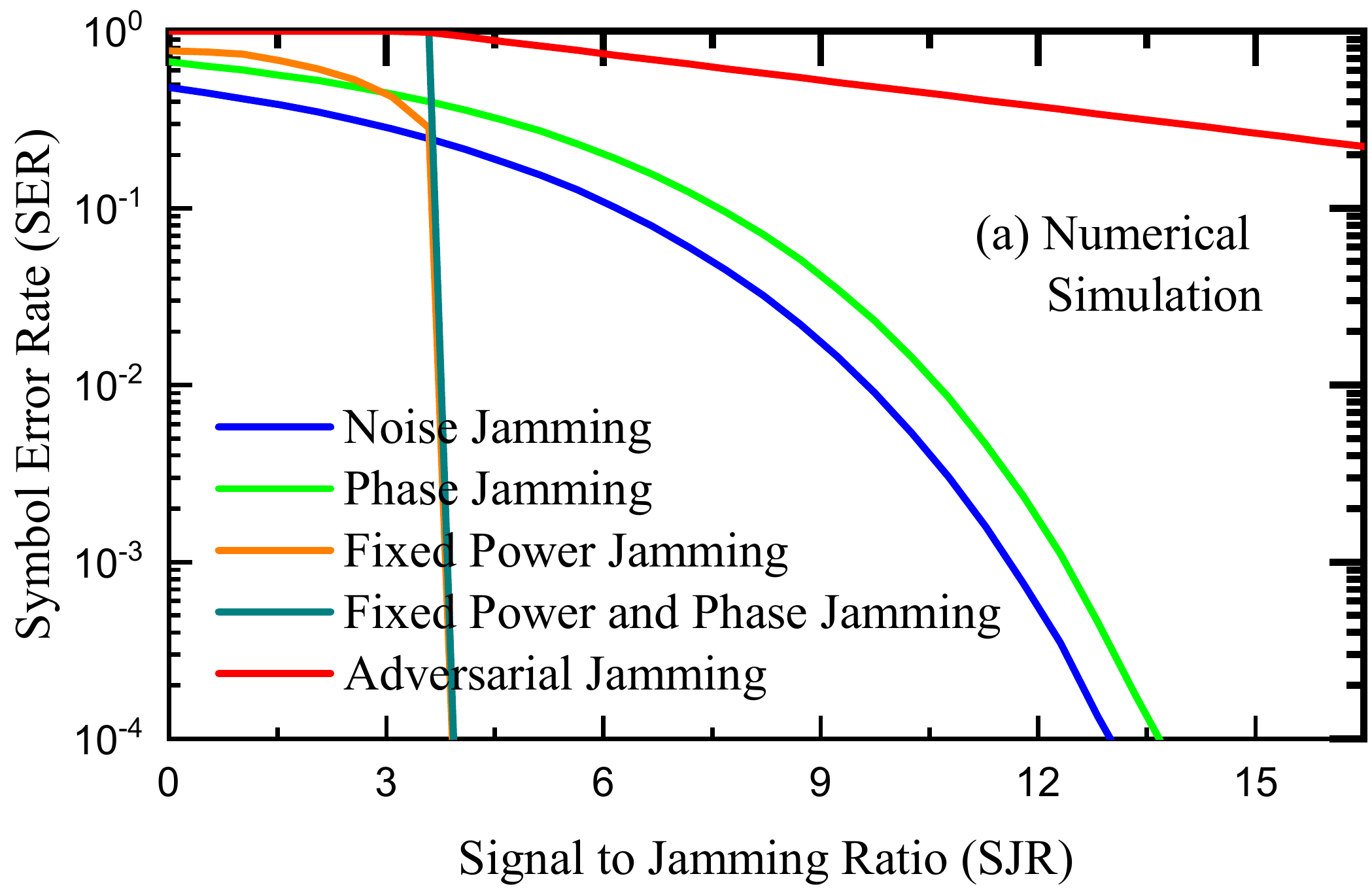}}
\centerline{\includegraphics[width=0.9\linewidth]{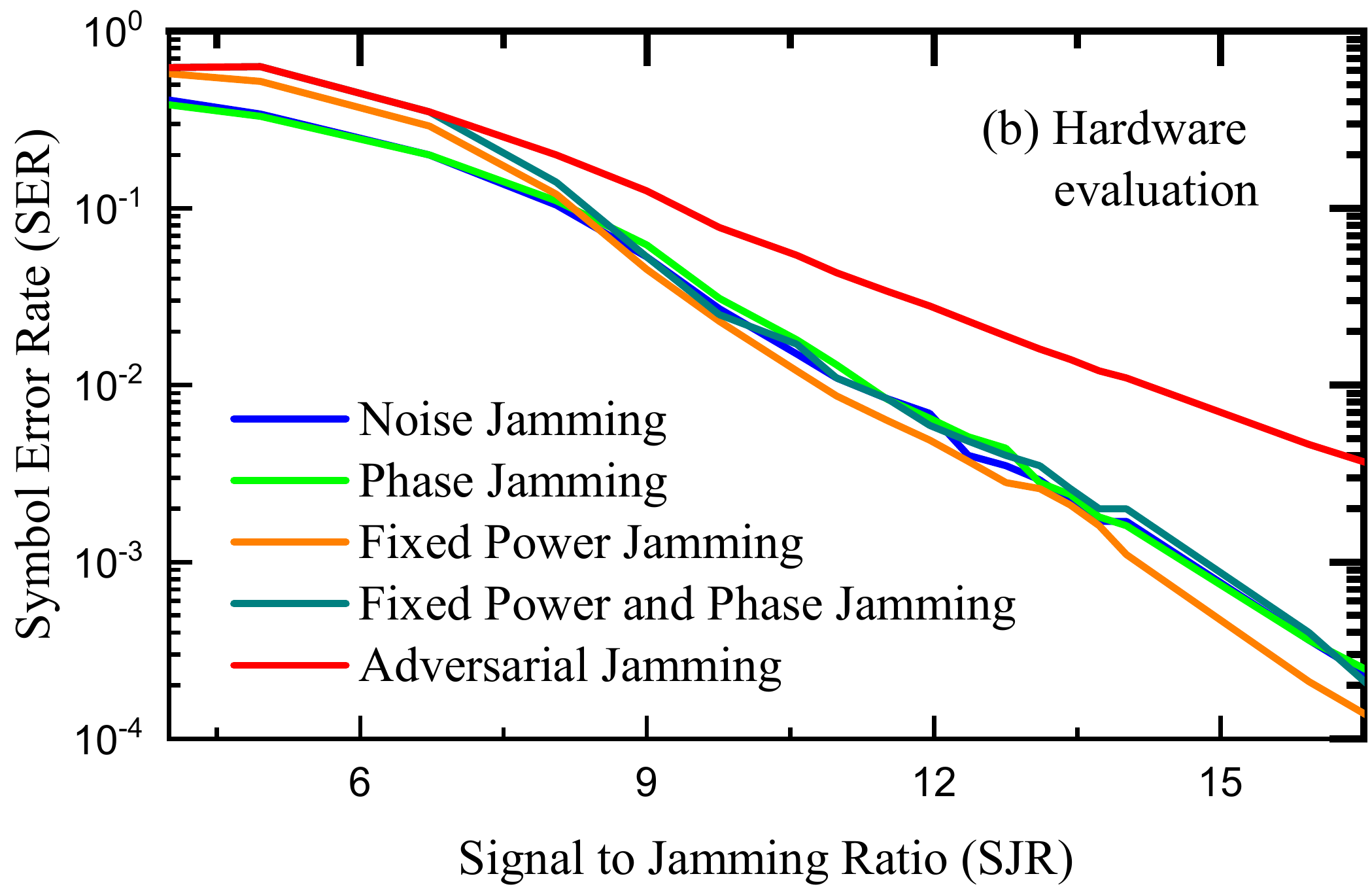}}
\caption{Comparison of the jamming effects of different jamming strategies numerically~(a) and by hardware~(b); AJ is the best.
\vspace{-5mm}
}
\label{fig:3}
\end{figure}

We take 16QAM as an example to show the impact of multiple jamming on SER under different SJRs, as shown in Fig. \ref{fig:3}(a). Each data averages a sequence of $500000$ bits.
On the whole, AJ shows the best jamming effect due to its waveform jamming ability with the optimal amplitude, phase, and time.

In traditional \textbf{noise jamming}, the SER curve shows a gentle downward trend as a whole. 
Hence, if the direction idea in AJ is introduced, that is, the phase of the jamming waveform is fixed according to the constellation diagram, but the amplitude is still arranged according to a Gaussian distribution, then the \textbf{phase jamming} curve can be obtained. Compared with that of a noise attack, the overall curve of a phase attack is slightly improved.

In contrast, if we keep the phase uniform distribution unchanged and use a jamming waveform with a fixed amplitude instead of a Gaussian distribution, we can obtain \textbf{fixed power jamming}. Obviously, the curve of this jamming is no longer a smooth and slowly decreasing curve; it has mutation points. A mutation point means that the jamming amplitude can confuse waveforms, so a jamming effect that is less than or greater than this JSR is very different.

Furthermore, the reduced time occurrence strategy is introduced and the phase is optimized to obtain \textbf{AJ}.
We pay attention to the AJ waveform and find that the modified waveform can be further improved on the basis of fixed power jamming.
Then, AJ can select the best jamming phase, estimate the required minimum jamming power, and ensure that the jamming waveform can have the expected jamming effect as much as possible.

\subsection{Hardware Testing}

In the field of signal research, numerical simulations can be used to verify the idea of a method, but hardware testing is essential because there will inevitably be unexpected factors. Therefore, we use NI-USRP-2954~\cite{USRP,UHD} equipment to restore the signal sending and receiving process to test and evaluate the performance of AJ.

We show the results of different jamming approaches in the hardware test in Fig.~\ref{fig:3}(b). On the whole, AJ has the best jamming performance, which is consistent with the numerical results.
The most significant difference between hardware testing and numerical testing is the introduction of clock synchronization because a receiver cannot obtain the starting time of each clock cycle of a waveform. Therefore, when the jamming is strong, synchronization becomes relatively difficult, and the overall SER result is larger than that obtained in the numerical simulation.

\subsection{Deception Attack}

Furthermore, we mentioned that a targeted attack can be realized by AEs, that is, a waveform can be disguised in a specified coded waveform, which is a signal deception attack.
The specific process is to generate the jamming waveform of target attack AEs so that the demodulator has error code behavior within the designed purpose.
It can be seen from the experimental results that we can exchange the waveforms of code `1100' and code `1000' without affecting other demodulation to achieve the goal of deception.

\section{Conclusion and Discussion}

Aiming at the problem of optimal jamming, an AJ waveform is proposed in this paper. The method is based on the idea of AEs, and the optimal amplitude, phase, and time of the jamming waveform are found during the optimization process. The best waveform jamming is realized, and the outstanding performance of the method is tested on numerical and hardware platforms. 
Our method has practical value and can realize deception attacks.
Finally, the purpose of this work is to jam an illegal spectrum, but there is the possibility of abuse, which requires the academic community to jointly improve the legal system.

\section*{Acknowledgment}

This work was supported by the National Natural Science Foundation of China (Grant No. 12004422) and by Beijing Nova Program of Science and Technology (Grant No. Z191100001119129).

\bibliographystyle{ieeetr}
\bibliography{LIW,paper_ref,AJ}

\begin{thebibliography}{10}

\bibitem{1637931}
W.~Xu, K.~Ma, W.~Trappe, and Y.~Zhang, ``Jamming sensor networks: attack and
  defense strategies,'' {\em IEEE Network}, vol.~20, no.~3, pp.~41--47, 2006.

\bibitem{5751298}
Y.-S. Shiu, S.~Y. Chang, H.-C. Wu, S.~C.-H. Huang, and H.-H. Chen, ``Physical
  layer security in wireless networks: a tutorial,'' {\em IEEE Wireless
  Communications}, vol.~18, no.~2, pp.~66--74, 2011.

\bibitem{871393}
D.~Pauluzzi and N.~Beaulieu, ``A comparison of snr estimation techniques for
  the awgn channel,'' {\em IEEE Transactions on Communications}, vol.~48,
  no.~10, pp.~1681--1691, 2000.

\bibitem{1512123}
K.~Baddour and N.~Beaulieu, ``Autoregressive modeling for fading channel
  simulation,'' {\em IEEE Transactions on Wireless Communications}, vol.~4,
  no.~4, pp.~1650--1662, 2005.

\bibitem{szegedy2013intriguing}
C.~Szegedy, W.~Zaremba, I.~Sutskever, J.~Bruna, D.~Erhan, I.~Goodfellow, and
  R.~Fergus, ``Intriguing properties of neural networks,'' in {\em ICLR}, 2014.

\bibitem{2014arXiv1412.6572G}
I.~J. Goodfellow, J.~Shlens, and C.~Szegedy, ``Explaining and harnessing
  adversarial examples,'' in {\em ICLR}, 2015.

\bibitem{jagannath_machine_2019}
J.~Jagannath, N.~Polosky, A.~Jagannath, F.~Restuccia, and T.~Melodia, ``Machine
  learning for wireless communications in the internet of things: A
  comprehensive survey,'' {\em Ad Hoc Networks}, vol.~93, p.~101913.

\bibitem{de_vrieze_cooperative_2018}
C.~de~Vrieze, S.~Barratt, D.~Tsai, and A.~Sahai, ``Cooperative multi-agent
  reinforcement learning for low-level wireless communication,'' {\em
  {arXiv}:1801.04541}.

\bibitem{madry2017towards}
A.~Madry, A.~Makelov, L.~Schmidt, D.~Tsipras, and A.~Vladu, ``Towards deep
  learning models resistant to adversarial attacks,'' {\em arXiv:1706.06083},
  2017.

\bibitem{cw}
N.~Carlini and D.~Wagner, ``Towards evaluating the robustness of neural
  networks,'' in {\em 2017 IEEE Symposium on Security and Privacy (SP)},
  pp.~39--57, May 2017.

\bibitem{USRP}
``{USRP X310}.'' {https://kb.ettus.com/X300/X310}.

\bibitem{UHD}
``{UHD}.'' {https://kb.ettus.com/UHD}.

\end{thebibliography}

\end{document}